\documentclass[twocolumn]{aastex7}
\usepackage{amsmath}
\usepackage{enumitem}
\usepackage{lineno}
\usepackage{float}
\usepackage{soul}
\DeclareMathOperator{\csch}{csch}
\usepackage{xspace} 
\usepackage{float} 
\usepackage{listings} 
\usepackage[skip=10pt plus1pt, indent=20pt]{parskip} 
\newcommand{\Msun}{\ \rm M_{\odot}}
\definecolor{darkgreen}{RGB}{0,128,0}
\definecolor{darkblue}{RGB}{0,0,255}
\definecolor{darkcyan}{RGB}{0,139,139}
\definecolor{darkbrown}{RGB}{139,69,19}
\definecolor{darkgray}{RGB}{169,169,169}
\definecolor{blacktext}{RGB}{0,0,0}
\definecolor{ForestGreen}{RGB}{34,200,34}

\lstdefinestyle{mystyle}{
commentstyle=\color{darkgreen},
keywordstyle=\color{darkblue}\bfseries,
numberstyle=\tiny\color{darkgray},
stringstyle=\color{darkbrown},
identifierstyle=\color{black},
basicstyle=\ttfamily\footnotesize\color{blacktext},
breakatwhitespace=false,
breaklines=true,
captionpos=b,
keepspaces=true,
numbers=left,
numbersep=5pt,
showspaces=false,
showstringspaces=false,
showtabs=false,
tabsize=2
}

\begin{document}
\linenumbers 
\title{Disentangling Drivers of Disk Warps in Tilted and Tumbling TNG50 Halos}
\author[0009-0002-3852-2034]{Saarthak Johri}
\email{sjohri@umich.edu}
\affiliation{University of Michigan Department of Astronomy 1085 S. University Ann Arbor, MI 48109, USA}

\author[0009-0003-7613-3109]{Neil Ash}
\email{nfash@umich.edu}
\affiliation{University of Michigan Department of Astronomy \\
1085 S. University \\
Ann Arbor, MI 48109, USA}

\author[0000-0002-6257-2341]{Monica Valluri}
\email{mvalluri@umich.edu}
\affiliation{University of Michigan Department of Astronomy \\
1085 S. University \\
Ann Arbor, MI 48109, USA}

\date{\today}

\begin{abstract}

\noindent 
Dark matter (DM) halos in $\Lambda$ Cold DM cosmological simulations are triaxial. Most exhibit figure rotation. We study 40 isolated halos with stellar disks from the TNG50 simulation suite across $\sim 4$~Gyr to understand whether and how a triaxial halo's tumbling and orientation relative to the disk can drive warps. We measure a warp angle $\psi$ and find even our isolated disks are all at least slightly warped, with each galaxy's maximum $\psi > 1.8^{\circ}$. We perform a modified cross-correlation analysis between $\psi$ and the figure rotation pattern speed, as well as the misalignment between the disk spin axis and (a) the figure rotation axis, (b) the halo minor axis, and (c) the gas angular momentum axis. We  use snapshots spanning a lookback time $t_{lb} ~4$ Gyr with 25 linearly-spaced lags from $ 0 - 2.33$ Gyr. We do not find evidence for a consistent lag between the onset of a warp and any of the aforementioned factors on the population level. However, we find significant correlations between individual time-series at various lags. These maximum correlation coefficients were significantly offset from random chance at the population level, suggesting that several of these factors do correlate with disk warping in specific situations. By examining four case studies whose maximum correlation coefficients were significantly higher than random chance, we establish clear qualitative relationships between these factors and warps. While a non-warped galaxy typically shows minimal halo tilt and figure rotation, warped galaxies can have strong/weak tilts and/or strong/weak figure rotation.

\end{abstract}

\section{Introduction} \label{sec: intro}

In the hierarchical model of DM (DM) halo formation within the $\Lambda$ Cold DM ($\Lambda$CDM) paradigm, angular momentum is imparted to DM halos during mergers and from large-scale tidal torques \citep{1992ApJ...401..441D}. There are two ways this angular momentum can manifest in a DM halo: first in DM particle streaming motions within the halo \citep{2009SSRv..148...37S}, and second (in non-spherical DM halos) in coherent tumbling\mdash similar to that of bars in stellar disks\mdash in a motion termed ``figure rotation.'' This investigation follows the study of \cite{ash2023figurerotationillustristnghalos} (hereafter AV23), which identified a catalog of DM halos which all exhibited steady figure rotation over several snapshots in the TNG50 simulation suite \citep{nelson2019ComAC...6....2N}. AV23 differs from previous studies of figure rotation of DM halos over the last few decades \citep[e.g.][]{1992ApJ...401..441D, Bailin_2004, 2007MNRAS.380..657B}, by analyzing figure rotation in DM halos \textit{with and without} baryonic physics (hereafter DM+B and DM-only, respectively). They found that most halos ($\sim84\%$) showed steady rotation over durations less than  $\sim 4$~Gyr. 
AV23 also found that figure rotation is only slightly affected by baryonic feedback and rather than disrupting or prohibiting figure rotation, feedback may drive DM halos to tumble slightly faster, with rotation axes slightly better aligned with the halo minor axis.

Although commonly overlooked in studies of secular galactic evolution, figure rotation has been shown to influence tidal streams \citep{2021ApJ...910..150V} as well as induce spiral arm and stellar bar formation \citep{2002ApJ...574L..21B}. In addition, \citet{Dubinski_2009} showed that torques and Coriolis forces from a halo tumbling about its minor axis can induce warping in an embedded disk using a set of idealized simulations. Because figure rotation is predicted to occur commonly in isolated galaxies under $\Lambda$CDM, it is worthwhile exploring whether figure rotation can induce warping in a cosmological environment containing contributions from other (potentially dominant) warp drivers. If studies of warped, isolated disks in cosmological simulations show that figure rotation has a high probability of being implicated as the driver, then observed warps in similarly isolated disk galaxies may provide empirical evidence that DM halo figure rotation occurs and may provide clues to the nature of the DM particle. 

Of course, disk warping can also be induced by several other factors: including misaligned gas accretion \citep{1989MNRAS.237..785O}, fly-by interactions \citep{2021MNRAS.508..541P}, and DM halos that are tilted with respect to their stellar disks \citep[e.g.][]{1983IAUS..100..177T, 10.1093/mnras/234.4.873, 1991ApJ...376..467K}. \cite{han2023tilteddarkhaloscommon} recently showed (for the TNG50 simulations) that a misalignment between the disk angular momentum axis and halo minor axis (hereafter `halo tilt') can be both long-lived and induce disk warps. They observe in one TNG50 Milky Way analog the formation of a warp  $\lesssim 1$ Gyr after the dark halo developed a tilt. For this halo, they also observe an evolution of the halo tilt angle as the warp develops, suggesting the importance of either halo tilt, halo figure rotation, or both as potential drivers. In a companion paper, these authors \citep{Han2023NatAst_warp} studied disk orbits in a rigid and tilted DM halo potential and showed that warping begins to develop after approximately one circular period (measured in the region of the warp, roughly $\sim400$ Myr). 

In this work, we examine three factors that have previously been considered to result in disk warps: halo figure rotation, gas misalignment and halo tilt. Since we rely on the figure rotation data from AV23, our sample consists of galaxies that are relatively isolated over the last 4 Gyr.  Additionally, we have visually confirmed all warps exhibited in the snapshots available to us are integral-shaped ``S-type'' warps, implying the cause of the warp is a force exerted over the whole disk, unlike e.g., ram pressure stripping connected to ``U-type'' warps \citep{Zee_2022} or the asymmetric ``L-type'' warps that are likely caused by tidal interactions \citep{Semczuk_2020}. This study is the first\mdash to our knowledge\mdash to examine the causes of warps in isolated (simulated) galaxies. By closely examining the nuances of halo and disk motion, we attempt to disentangle the various dynamical processes associated with disk warps. One of the objectives of this study is to assess whether disk warps can be used to probe the figure rotation of DM halos.

The rest of this paper is outlined as follows: In Section \ref{sec: theory}, we describe the simulation suite and the selection of simulated halos used. In Section \ref{sec: methods}, we review the procedures for measuring halo shape, figure rotation, and describe how we measure the warp angle $\psi$. In addition, we describe how we measure the orientation of the halo principal axes, the disk spin axis, gas angular momentum, and the angular separation between these axes with their associated uncertainties. In Section \ref{sec: Results}, we present our analysis of the warps in the TNG50 disks and attempts to obtain quantitative correlations between the presence of a warp and its causes. We focus on four case studies that exhibit warps and show that $\psi$ is affected to varying degrees by halo tilt and figure rotation. We discuss the implications of these results and conclude by summarizing key findings in Section \ref{sec: summary}.

\section{Simulations} \label{sec: theory}

We use the TNG50 simulations, the highest-resolution realization of the IllustrisTNG cosmological simulation suite \citep{nelson2019ComAC...6....2N, nelson2019MNRAS.490.3234N, p2019MNRAS.490.3196P}. Using the cosmological magnetohydrodynamics code AREPO \citep{Springel_2010}, TNG50 simulates $2 \times 2160^3$ DM and gas particles in a comoving box of side length 50 Mpc\mdash resulting in a mass resolution of $4.5 \times 10^5 \Msun$ and spatial resolution of ~100-140 pc. TNG50 assumes the Planck 2015 cosmology \citep{2016} with $H_0 = 67.8$ kms$^{-1}$Mpc$^{-1}$, $\Omega_m = 0.308$, and $\Omega_bh^2 = 0.022$. The TNG50 simulations span  $z=127$ to $z=0$ (public snapshots run from $z=20$ to $z=0$), and we use the final 25 snapshots $0 \leq z \leq 0.35$ (3.966 Gyr). This range yields a mean temporal resolution of $107h^{-1}$ Myr, ranging from $80h^{-1}$ Myr to $160h^{-1}$ Myr. 

TNG50's halo and subhalo catalogs are generated, respectively, by the FOF \citep{2005Natur.435..629S} and SubFind \citep{2001MNRAS.328..726S} algorithms. The LHaloTree \citep{2015A&C....13...12N} merger tree\mdash generated using the FOF halos identified by Group Number (GrNr) \mdash tracks halos across snapshots. We make use of the halo catalog published in AV23, which is restricted to halos whose virial masses are between $10^{10}-10^{13}$ $h^{-1}\Msun$. To remove significant gravitational perturbations from outside sources, they used merger-eliminating criteria similar to those reported in \cite{Bailin_2004}, excluding halos whose total substructure mass is $>5\%$ of the virial mass of the primary halo or who accrete $>10\%$ of the primary between any two snapshots. This was necessary since figure rotation measurements that rely on halo shape measurements made using the shape tensor method are sensitive to the presence of massive subhalos. The AV23 catalog is cross-matched between the DM-only and DM+B run using the bi-directional subhalo matching catalog generated by \cite{nelson_illustris_2015}. After applying these criteria, their final catalog numbered 1,754 DM+B halos. Although AV23 applied further cuts on halos with poor shape measurements\mdash bringing the catalog of figure rotation measurements down to 1,396 DM+B halos\mdash we use the full sample of 1,754 DM+B halos.

\section{Methodology} \label{sec: methods}

\subsection{Disk Galaxy Identification} \label{sec: disks}

We initially identified the halos containing stellar disks by using an IllustrisTNG supplemental data catalog consisting of the circularities, angular momenta, and axis ratios of stellar components for a sub-sample of TNG50 halos \citep{Genel_2015} (hereafter G15). After rotating the coordinate system such that the z-axis is aligned with the total angular momentum of all stellar particles within 10 times the stellar half mass radius ($10r_{0.5}$, G15 calculated the circularity parameter $\epsilon$ for each star as
\begin{equation}
    \epsilon = \frac{j_z}{\boldsymbol{|j_\mathrm{circ}}(E)|},
\label{eq:epsilon}
\end{equation}
where the $z$-component of each stellar particle's specific angular momentum ${j_z}$ is divided by the absolute value of the specific angular momentum  of a circular orbit with same energy as the particle $|\boldsymbol{j_\mathrm{circ}}(E)|$. They reported the fractional mass $F_{*,\mathrm{disky}}$ of stars with $\epsilon > 0.7$. Disks are assumed to have $F_{*,\mathrm{disky}} > 0.5$. Since G15 imposed a  mass cut ($M_* > 3.4 \times 10^8 \Msun$), their supplementary catalog and the catalog of 1,754 DM+B halos from AV23 have only 550 halos in common, resulting in only 23 galaxies which were disky over all snapshots (75-99).

To increase the sample of disks and include the galaxies with lower stellar mass in the sample of AV23, we developed an alternative algorithm for computing  $\epsilon$ and the fraction of star particles  with $\epsilon > 0.7$. 
After aligning the coordinate system with the total angular momentum of all stellar particles within $r_{0.5}$ (as to not bias the measurement with warping at the outer edges), we estimate the gravitational potential of the galaxy from masses of all DM, star, and gas particles lying within $10r_{0.5}$. The potential is estimated assuming axisymmetry, with the symmetry axis defined by the angular momentum axis of stars with $r<r_{0.5}$. This oblate axisymmetric potential is computed using a basis function expansion (BFE) algorithm within the AGAMA galactic dynamics toolbox \citep{2019MNRAS.482.1525V}, specifically a multipole expansion ($m = 0$, $l_\mathrm{max}=8$, $r_\mathrm{min}=0.3$ $h^{-1}$kpc, $r_\mathrm{max}=10r_{0.5}$, and $N_\mathrm{grid}=25$). We then determine $\epsilon$ for each particle with Equation~\ref{eq:epsilon} using the total binding energy of each particle $E$, the specific angular momentum of the circular orbit with energy $E$ ($\boldsymbol{j_\mathrm{circ}}(E)$).

\begin{figure}
\centering
\includegraphics[trim=7pt 7pt 6pt 6pt, clip, width=0.95\columnwidth]{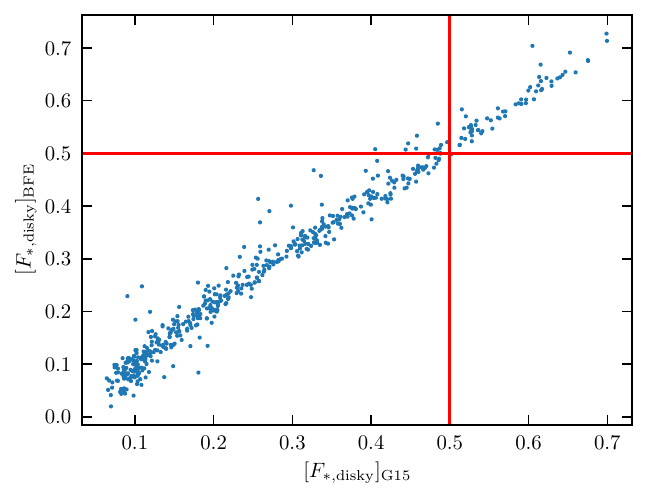}
\caption{Fractional mass of disky stars $F_{*,\mathrm{disky}}$ from our BFE-based measurement vs. the estimate of G15 for each of the 550 galaxies that overlap between the catalogs of G15 and AV23.  The slope of the correlation is fairly close to unity. G15 considered all galaxies with $F_{*,disky} > 0.5$ to be disks (vertical red line), and we impose the same restriction (horizontal red line). The galaxies in the rectangular region at the top right of the plot are considered disks by both methods. 
}
\label{fig:g15}
\end{figure}

To test the validity of our measurements against those of G15, we calculated $F_{*,\mathrm{disky}}$ for the 550 galaxies in common to samples of AV23 and G15. Figure \ref{fig:g15} shows a tight correlation between our BFE-based measurements of $F_{*,\mathrm{disky}}$ ($\mathrm{y}$-axis) and the G15 measurements, with our estimates being slightly stricter for each galaxy. All disky galaxies with $F_{*,\mathrm{disky}}>0.5$ (right of the vertical line) by the G15 criterion would also be considered disks by our method, save a single galaxy that doesn't pass our criterion (below the horizontal line). By applying this BFE-based method to the galaxies excluded by G15's mass cut, we expanded our sample to 40 DM+B halos with disks in the stellar mass range [$10^8 \Msun,10^{10} \Msun$], measured within the usual $2r_{0.5}$. 

\subsection{Halo Shape Measurements and Uncertainties} \label{sec: shape}

AV23 measured the DM halo shapes and orientations of the principal axes by iterating over the shape tensor. They quantified halo shapes with new ellipticity and prolateness metrics $e$ and $p$:
\begin{equation}
    e = 1 - \frac{c}{a};  \; p = 1-2\frac{b}{a}+\frac{c}{a},
\end{equation}
where ${c}/{a}$ and ${b}/{a}$ are the axis ratios of the minor:major and intermediate:major axis lengths, respectively.

Two factors limit our ability to measure the axis lengths: the halo's shape and the particle number $N_\mathrm{part}$. The uncertainty on halo shape measurements is Poisson-noise dominated and hence scales inversely with $\sqrt{N_\mathrm{part}}$ \citep[e.g.][]{Bailin_2004,ash2023figurerotationillustristnghalos}. We adopt the semi-analytic axis uncertainty prescriptions from AV23. These were generated by performing least-squares fitting on uncertainty estimates from jackknife-resampled halos within the AV23 catalog described in Section \ref{sec: theory}. Those semi-analytic functional forms are the hyperbolic cosecant functions below:
{\setlength{\abovedisplayskip}{0pt}
\setlength{\belowdisplayskip}{0pt}
\begin{align*}
\sigma_x = 1.01 \times 10^{-2}\sqrt{N}\csch[8.17\times10^{-1}(1+\frac{p}{e})],\\ 
\sigma_z = 5.34\times10^{-6}\sqrt{N}\csch[5.84\times10^{-4}(1-\frac{p}{e})].
\end{align*}}
We only use and therefore report the analytical approximations for the orientation uncertainties on the major and minor axes, $\sigma_x$ and $\sigma_z$.

To make measurements of halo figure rotation, AV23 generalized the best-fit plane method of \mbox{\citep{Bailin_2004}} to measure the pattern speed ($\Omega_p$) of figure rotation about an arbitrary axis. We make use of the halo catalogs identified by AV23, but instead generate pattern speed measurements between successive snapshots using the quaternion method first described by \mbox{\cite{Bailin_2004}}. This approach allows us to probe figure rotation on timescales shorter than those considered by AV23, while retaining sensitivity to arbitrary rotation axis orientation and rotation duration.

\subsection{Angular Separation Calculations} \label{sec: angles}

We calculate the orientation of a disk's spin axis by computing angular momentum within $r_{0.5}$. As mentioned before, this restriction avoids biases due to the warp and/or low particle counts at the outer edge. We then calculate the angle between this disk spin axis and halo minor axis. The angle between the minor and disk spin axes are referred to hereafter as its tilt angle. To factor in gas accretion as a warp driver, we also calculate the angular momentum of the gas within $10r_{0.5}$ for every galaxy and calculate its misalignment with the angular momentum of the disk. We do not place a restriction on gas temperature as hot gas is either blown out of the galaxy and/or is not misaligned with significant angular momentum. The orientations of the disk spin axis as well as the halo minor and major axes are decomposed into azimuthal ($\phi$) and polar ($\theta$) components in various reference frames for our polar plot analysis in Section \ref{sec: Results}.

\begin{figure*}[th]
    \centering
\includegraphics[width=1.95\columnwidth]{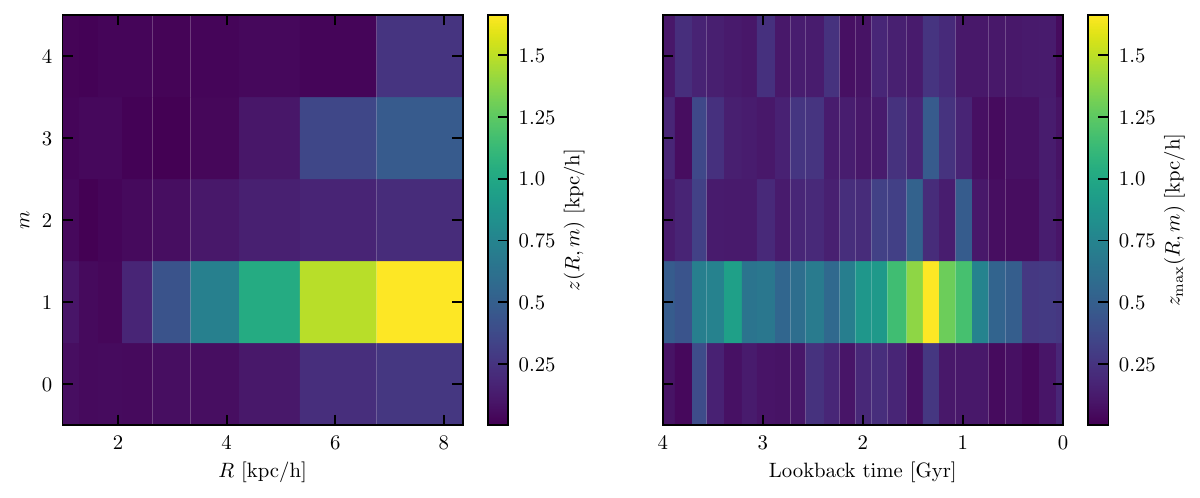}
\caption{Left: Different Fourier modes for Halo 1292 at Snapshot 91 (seen in Figure \ref{fig:warp fitting}) as a function of radius with the color signifying $z(R,m)$ (Eq. \ref{eq:warp}). The only significant mode is $m = 1$. Right: The maximum Fourier amplitude over all radii as a function of lookback time, again showing that the only significant mode is $m=1$.}
\label{fig:mode}
\end{figure*}

\subsection{Warp Measurements and Uncertainties} \label{sec: warp}

To measure the warp of the disk, we first define a cylindrical coordinate system with $R$ the cylindrical radius and the $z$-axis parallel to the disk spin axis. Inspired by the azimuthal harmonic density decomposition commonly used to measure bar strength, we perform an azimuthal harmonic decomposition of the disk height above/below the inner disk plane. Specifically, in a series of radial bins we measure the harmonic amplitude: 
\begin{equation}
    z(R,m) = \frac{2}{\sum_j M_j} \left| \sum_j M_j z_j \exp(i m\phi_j) \right|,
\label{eq:warp}    
\end{equation}
where each particle has azimuthal angle $\phi_j$, vertical height $z_j$ and mass $M_j$. The harmonic amplitude $z(R,m)$ is hereafter referred to as the warp amplitude. While not commonly appearing in the literature, this method has been used at least once before by \cite{Dubinski_2009}, and is advantageous in that it does not assume any particular radial warp profile.

\begin{figure}
\centering
\includegraphics[trim=7pt 7pt 6pt 6pt, clip, width=0.95\columnwidth]{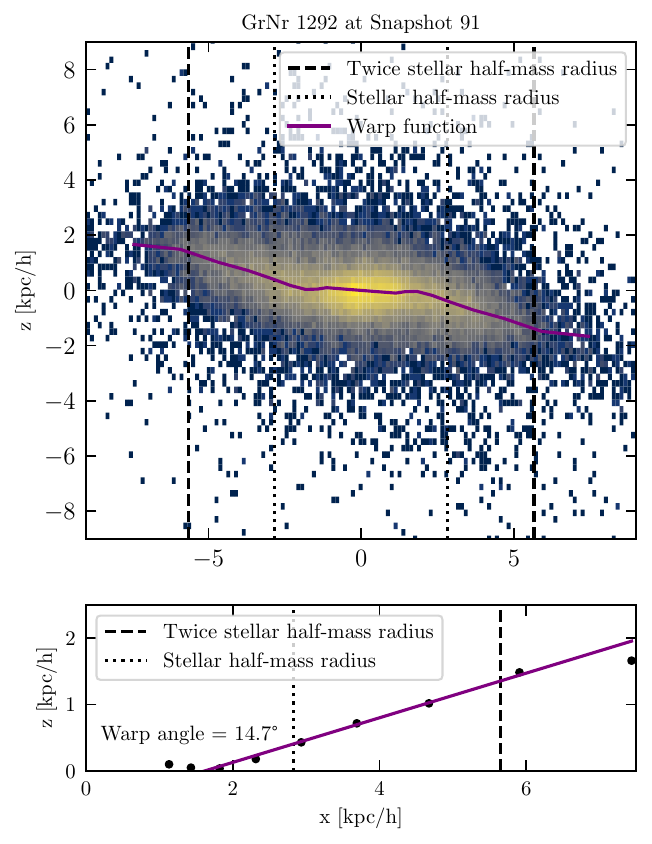}
\caption{Top: Warp function (Eq. \ref{eq:warp}) reflected and plotted over 2-D histogram of our most warped disk. The vertical dotted lines and the vertical dashed lines mark $2r_{0.5}$ and $r_{0.5}$ on the left and right of the galaxy's center, respectively.
Bottom: Linear fit of the above warp function in the $r_{0.5}$ to $2r_{0.5}$ region, with the angle of the line annotated.}
\label{fig:warp fitting}
\end{figure}

For our purposes, the fundamental mode ($m=1$) suffices, as demonstrated by Figure \ref{fig:mode}. The in-plane warp phase-angle $\varphi$ is measured in a given radial bin as
\begin{equation}
\varphi(R) = \frac{1}{m}\arctan\left.\left(\frac{\sum_jM_jz_j \sin(m\phi_j)}{\sum_jM_jz_j \cos(m\phi_j)}\right)\right|_{m=1},
\end{equation}
and is used to appropriately rotate our fitting over an image of the warped galaxy (as seen in the top panel of Figure \ref{fig:warp fitting}. 

We measure the warp angle $\psi$ by fitting a line to $z(R,1)$ in Eq. \ref{eq:warp} as a function of radius in the region between $r_{0.5}$ and $2r_{0.5}$, as seen in the bottom panel of Figure \ref{fig:warp fitting}. In this region,  $z(R,1)$ is broadly linear. We favor this measurement over previous studies that define the warp amplitude as the $z$-height of a disk at a given radius (e.g. twice the half mass radius,  $2r_{0.5}$) due to the fact that warps in many massive and/or elongated galaxies in our sample can extend beyond $2r_{0.5}$.

Uncertainties are generated on all calculated angular measurements by running the measurement processes through 100 bootstrap realizations, each sampling 90$\%$ of the full population of disk stars with replacement.

\section{Results} \label{sec: Results}

\subsection{Warp Distribution} \label{sec: dist}

\begin{figure}
\centering
\includegraphics[trim=7pt 7pt 6pt 6pt, clip, width=0.95\columnwidth]{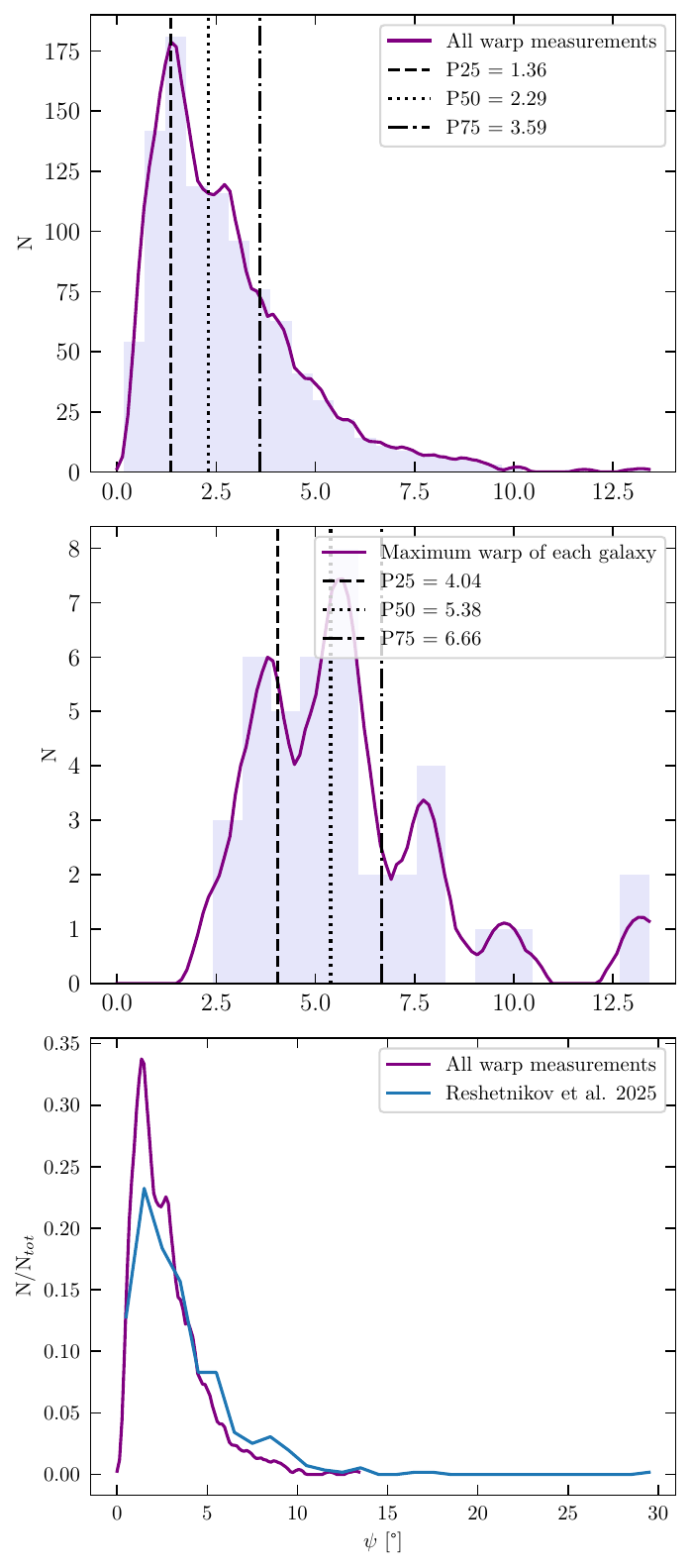}
\caption{KDE of $\psi$ measurements for all 40 galaxies across 25 snapshots (top) and maximum $\psi$ of each of the 40 galaxies (middle). The vertical lines show the 25th, 50th and 75th percentile intervals. The bottom panel shows that the fractional distribution of $\psi$ measurements for our simulated sample (magenta curve) is remarkably similar to the distribution for $\sim$1000 observed edge-on galaxies with 'S-type' warps from \citet{refId0}.}
\label{fig:kde}
\end{figure}

We analyzed the evolution of warp angle $\psi$ over the last 25 snapshots ($\sim$4 Gyr) for the 40 disks in our sample. The top panel of Figure \ref{fig:kde} shows a kernel density estimate (KDE) of the distribution of $\psi$ measurements for all 40 disks at each of the 25 snapshots. While most $\psi$ values do not exceed 5 degrees, Halo 1292 (shown in Figure \ref{fig:warp fitting}) is very warped and contributes several outliers to the distribution. The distribution of the maximum $\psi$ for each of the 40 galaxies (Figure \ref{fig:kde}, middle panel) shows several distinct peaks, but it is unclear if these are significant due to the small numbers. The bottom panel of Figure \ref{fig:kde} compares our $\psi$ distribution from the top panel normalized by the total (magenta curve) with the distribution of observed S-type warps from \cite{refId0} for the largest available sample ($\sim$1000) of distant edge-on galaxies (blue curve). Their warp angles $\psi$ is measured in a similar way to ours and the similarity of the two distributions is remarkable. It is worth noting that every system in our sample of 40 halos whose maximum warp angle $\psi_{\max}$ fell above the median exhibited significant tilt and figure rotation. In the following subsections, we explore the possible correlations and co-evolution between $\psi$ and these quantities.

\subsection{Quantitative Cross-Correlation Analysis}
Our objectives in this work are  to determine the following. (1) Can the strength of previously proposed drivers  of warps (halo tilt, figure rotation, gas misalignment) be quantitatively correlated with warp angles at later times? (2) Is there evidence for a well defined time-lag between the onset of the perturbation induced by a given warp driver and the time required for a disk to respond\footnote{e.g., \citet{han2023tilteddarkhaloscommon} observed a time lag of order of 1~Gyr between the development of a halo tilt the onset of disk warping in their archetypal system.}? To address these questions, we attempt to measure correlations between $\psi$ at a given snapshot with the following quantities at some earlier time: halo tilt, misalignment between the figure rotation axis and disk angular momentum, figure rotation pattern speed, and misalignment between the momentum of gas within 10$r_{0.5}$ and disk angular momentum. 

\begin{figure*}
\centering
\includegraphics[width=1.95\columnwidth]{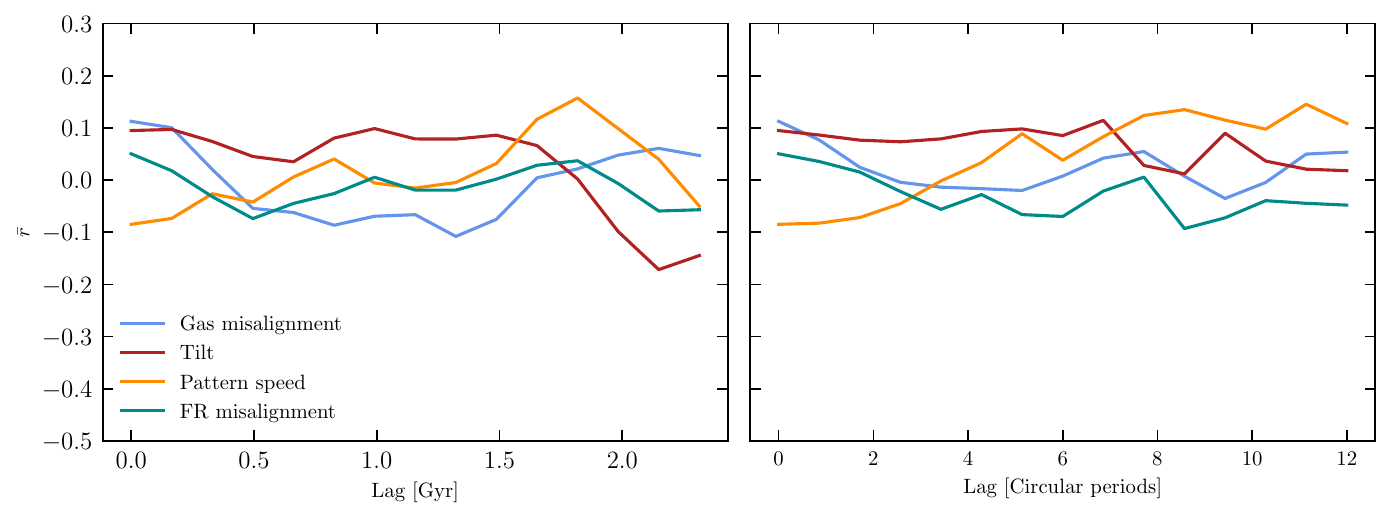}
\caption{Left: Averaged Pearson correlation coefficient $\bar{r}$ (from eq.~\ref{eq:rbar}) between warp angle $\psi$ and potential drivers as a function of time lag in Gyr (see text for details). Right: The same as the left plot with units in the circular periods of each galaxy calculated at the stellar half mass radius at Snapshot 99. None of the $\bar{r}$ values imply even moderate correlation, implying no uniform lag time over the sample. The p-values averaged over all calculations listed in the legend are also all too high to signify meaningful correlation.}
\label{fig:correlation}
\end{figure*}

We perform a modified cross-correlation analysis with the following steps. First, to account for the non-uniform chronological spacing between TNG's snapshots, we interpolate all measurements in time onto a uniformly-spaced grid of 25 points (hereafter `interpolated snapshot') using a cubic spline. We choose 25 interpolated snapshots to match the original number of snapshots available within the time series, and note that our interpolated snapshots have a fixed separation of $\sim0.167$ Gyr. For each given time lag, we then calculate the Pearson correlation coefficient $r$ between each of the aforementioned warp drivers and the warp angle $\psi$ a number of snapshots later, with the number determined by the time lag. The Pearson correlation for each lag time is measured using all available snapshot pairings separated by that lag time. For example, to test a lag of $\sim 0.5$ Gyr, we calculate $r$ between a warp driver at a given snapshot and $\psi$ three snapshots later (Snapshot 75 with Snapshot 78, Snapshot 76 with Snapshot 79, etc). In figure \ref{fig:correlation} we report $\bar{r}$, which is averaged over the 40 galaxies in our sample: 
\begin{equation}
\bar{r}(t_{\mathrm{lag}}) = \frac{1}{n_{\mathrm{gal}}} \sum^{n_{\mathrm{gal}}}_{i=1} r_i(t_{\mathrm{lag}})\label{eq:rbar},
\end{equation}
where $r_i(t_{\mathrm{lag}})$ is Pearson's correlation coefficient measured for galaxy $i$ using all interpolated snapshots separated in time by $t_{\mathrm{lag}}$, and $n_{\mathrm{gal}}=40$.

We repeat our correlation calculation for snapshot separations (lags) ranging from $\sim 0-2.33$ Gyr. We choose an upper lag limit of 2.33 Gyr for two reasons: 1) larger lags have fewer snapshot pairings between which we may measure correlations, possibly leading to numerically spurious results, and 2) a lag of $\sim2.3$ Gyr corresponds to $\gtrsim15$ circular periods in most of our disks, whereas disks should only require a few circular periods to respond to external perturbations. By placing an upper limit at $\sim2.3$ Gyr, we ensure that we stay in a physically relevant lag range and have at least 11 snapshot pairings available at any given lag. Because our interpolated snapshots have a spacing of 0.167 Gyr, we are able to test a total of 15 discrete time lags in our range of $0-2.33$ Gyr.

If a particular warp driver is responsible for many of the warps in our catalog and generates these warps on approximately the same timescale, then we expect to see a local maximum in $\bar{r}$ at the lag corresponding to the time needed for the warp to develop. Figure \ref{fig:correlation} plots $\bar{r}$ as a function of lag time for each of the four drivers being tested: halo tilt (red), figure rotation pattern speed (orange), figure rotation misalignment (green), gas misalignment (blue). In the left panel, we show Pearson's $r$ correlation coefficient averaged over all 40 galaxies at each of our 15 fixed, physical lag times. The mean correlation coefficient $\bar{r}$ exceeds 0.15 only for the pattern speed of halo figure rotation, making it the best correlated quantity among the population averages. Although the $\bar{r}$ curves show some time evolution with local maxima present, the overall low $\bar{r}$ values do not suggest statistically significant correlations and therefore do not support a preferred lag time for any of our four warp drivers.

The apparent lack of correlation or a common lag between each of our possible warp drivers and disk warping when averaged over galaxies using our physical time lags could be caused by different dynamical times of the galaxies in our sample. We modify our averaging scheme in eq. \ref{eq:rbar} to account for our galaxy's individual circular periods by first constructing cubic splines of $r_i(t_{\mathrm{lag}})$ for each of our 40 galaxies. We then sample each interpolated $r_i$ at different multiples of the galaxy's circular period. The circular periods are measured at present day\footnote{We assume that each galaxy's circular period does not evolve significantly over the last 4 Gyr.} using the total mass enclosed by the half-mass radius and assuming spherical symmetry. We choose a set of 15 `normalized' time lags spaced linearly between 0 and 12 circular periods on which to measure the averaged correlation coefficient $\bar{r}(t_{\mathrm{lag}})$, and show these averages in the right panel of Figure \ref{fig:correlation}. Accounting for the disparate circular periods in our galaxy sample evidently does not improve our averaged correlation coefficients; the amplitudes of $\bar{r}$ remain relatively unchanged, and rather than becoming more sharply peaked about any lag time instead show little variation over lags spanning $0-12$ circular periods for any of our 4 warp drivers. We therefore do not find a particular lag time, either in physical time or dynamical times, that yields a strong correlation between the warp angle and any of its potential drivers for the average galaxy in our sample.

\begin{figure*}
    \centering
\includegraphics[width=1.95\columnwidth]{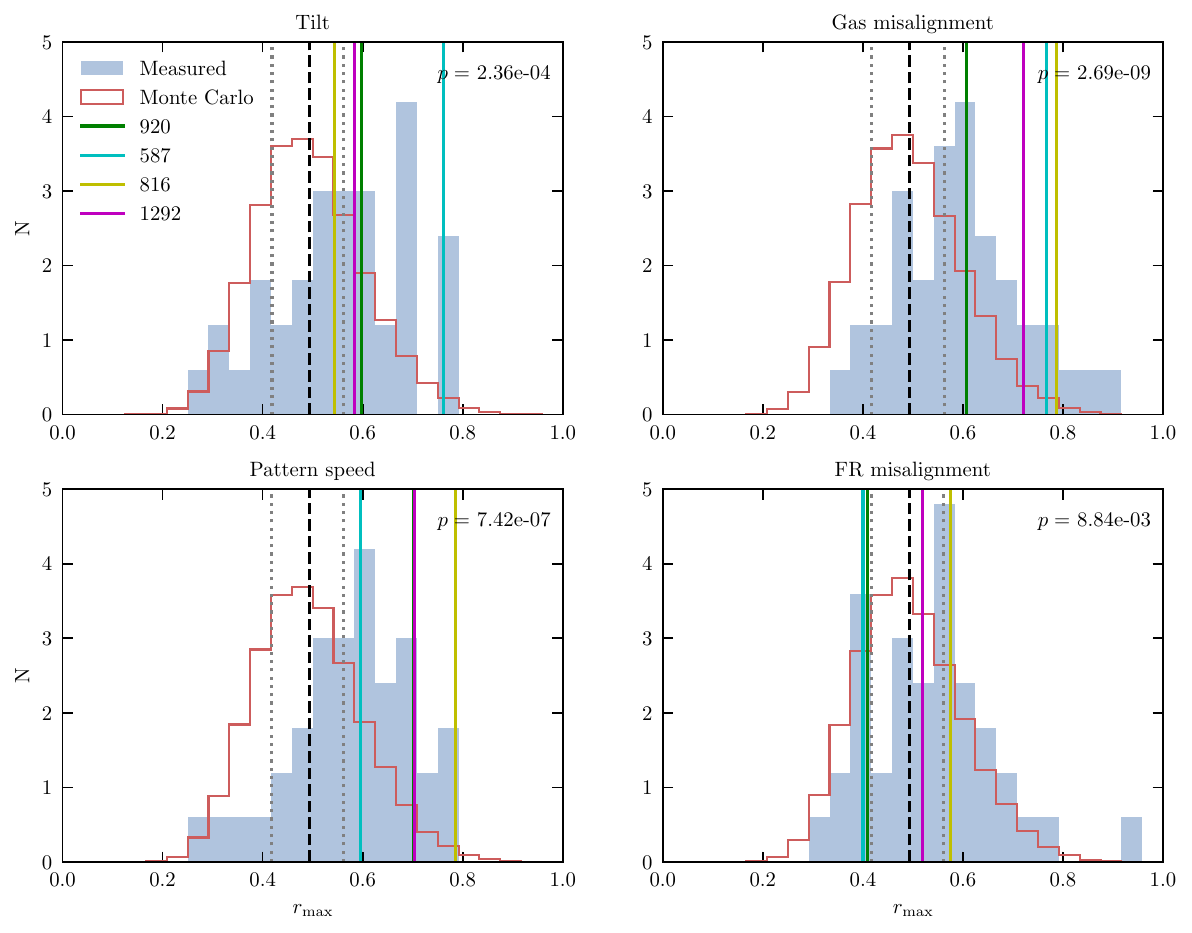}
\caption{The distribution of the maximum Pearson $r$ statistic $r_{\mathrm{max}}$ achieved by each galaxy in our lag analysis compared to the same analysis conducted on random (Monte Carlo) data drawn from Gaussian distributions with means and variances matching those of the observed time series. We perform Student $t-$tests between the observed $r_{\max}$ distributions and the Monte Carlo distributions to assess how consistent our measured correlations are with random chance, and display the associated $p-$values in the top right of each panel. Our measured $r_{\max}$ distributions are significantly offset to higher values compared to the Monte Carlo distribution at the population level, suggesting that each of our four warp drivers may contribute to disk warping in a typical system. The dashed black, dashed gray, and solid colored vertical lines indicate mean, standard deviation, and case study $r_{\mathrm{max}}$, respectively. We selected case studies based on isolated, qualitatively correlated behaviors and for achieving $r_{\mathrm{max}}$ at least 2$\sigma$ from the Monte Carlo mean, suggesting that their correlations are not likely to have been achieved by random chance. FoF GrNr's 920, 587, 816, and 1292 show significant correlations for pattern speed, tilt and gas misalignment, pattern speed and gas misalignment, then finally pattern speed and gas misalignment respectively.}
\label{fig:MCMC}
\end{figure*}

We then zero in on the maximum correlation coefficient achieved by each galaxy's cross-correlation (hereafter $r_{\max}$) as it is a potentially informative metric that could encode how important a given warp driver is for inducing subsequent warping. However, the statistical significance of a given $r_{\max}$ is ambiguous. To assess whether the $r_{\max}$ we measure are consistent with random uncorrelated data, we perform a series of Monte Carlo (MC) experiments. In each experiment, we sample 25 random, uncorrelated mock data points from Gaussian distributions whose means and variances match the observed means and variances of our interpolated time series. We then repeat our full cross-correlation analysis on the random MC data to measure the resultant $r_{\max}$. For each warp driver and each galaxy, we perform 1000 MC experiments to approximate the distribution of $r_{\max}$ achieved by random chance from uncorrelated data. 

The MC $r_{\max}$ distributions (averaged over all galaxies) are compared to our observed $r_{\max}$ distributions in Figure \ref{fig:MCMC}. For each of our 4 warp drivers, we perform a Student $t$-test between the two distributions to assess whether the observed $r_{\max}$ distributions are consistent with random chance. $p-$values for this test are reported in the top right corner of each panel in Figure \ref{fig:MCMC}. The distributions for gas misalignment and figure rotation pattern speed $r_{\max}$ show the largest offset from our MC distributions, with $t-$test $p-$values of $\sim3\times10^{-9}$ and $\sim7\times10^{-7}$, respectively. The figure rotation misalignment $r_{\max}$ distribution shows the weakest offset from random chance, suggesting that this offset may not be a significant driver of disk warping in our sample of 40 cosmological disks. While the evidence is somewhat weak in the case of figure rotation misalignment, we observe an offset towards higher $r_{\max}$ compared to random chance at the population level for each of our four warp drivers, which suggests that each  may contribute to disk warping in our sample.

While our population $r_{\max}$ values are shifted significantly with respect to the MC $r_{\max}$ distributions, it is apparent in Figure \ref{fig:MCMC} that many of the galaxies in our sample have $r_{\max}$ values which are close to or below the MC mean. This suggests that while each of our drivers may contribute to disk warping in a significant way at the population level, they may not necessarily contribute significantly in an individual galaxy (which could happen if, for instance, the torques due to accreting gas are so large as to make the torques from a tilted halo inconsequential). For building intuition, we select four of our disk galaxies which each show $r_{\max}$ values $\geq2\sigma$ above the MC mean for 1 or more of our warp drivers (shown in Figure \ref{fig:MCMC}) and look at the detailed evolution of their halos and disks. 

\subsection{Case Studies} \label{sec: archetypes}

We identified four case studies (one without a warp and three with warps) from our sample of 40 disk galaxies. They show a range of correlations over time between the (a lack of) warp angle and halo properties such as tilt angle relative to the disk, the absence/presence of halo figure rotation, as well as the misalignment between disk and gas angular momentum axes. These case studies were selected because they were outliers in the distribution of $r_\mathrm{max}$ for one or more of the warp drivers (see Figure~\ref{fig:MCMC}).

\begin{figure*}
\centering	%
    \includegraphics[trim=7pt 7pt 6pt 6pt, clip, width=0.8\columnwidth]{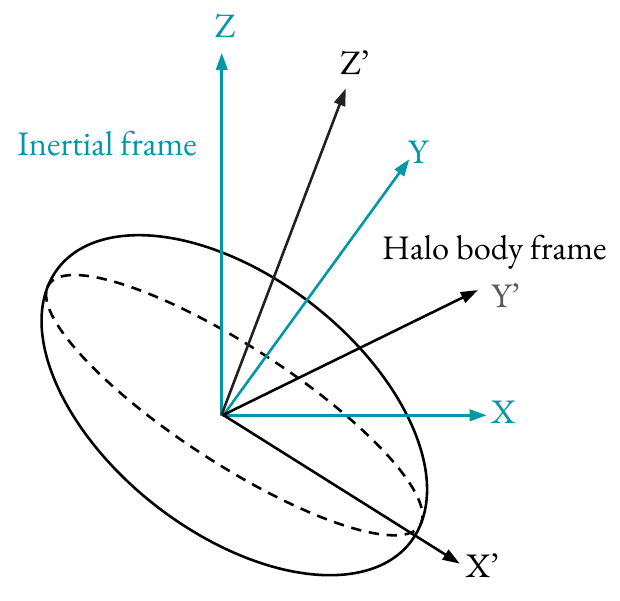}
	\includegraphics[trim=4pt 4pt 3pt 3pt, clip, width=0.8\columnwidth]{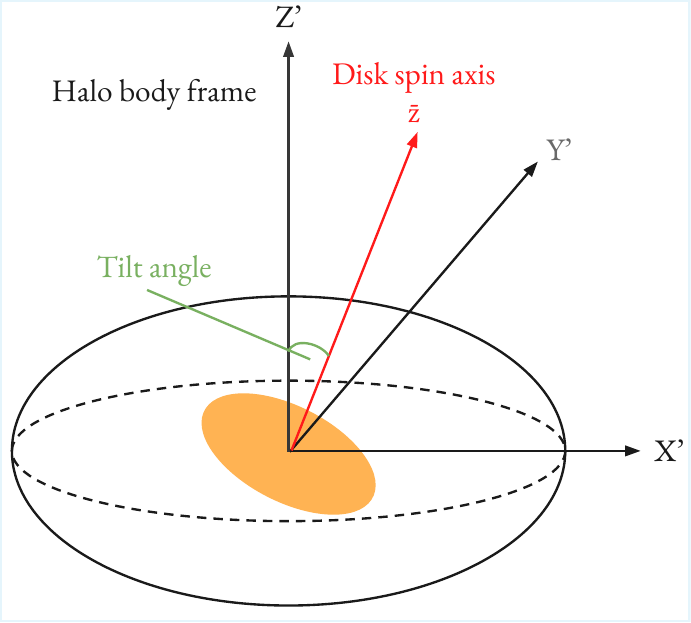}
\caption{Left: Illustration of the relative orientations of the halo body frame coordinate system (black $X', Y', Z'$) to the simulation inertial frame (teal $X, Y, Z$). In  Figs.~\ref{fig:920}-\ref{fig:1292} solid arrows represent the various frames: the halo body frame (black: $X', Y'$), inertial frame (teal $X, Y$). Right: Tilt of the disk spin axis (red $\bar{z}$) relative to the halo body frame, with the tilt angle labeled.}
\label{fig:frames}
    \centering
\includegraphics[width=1.95\columnwidth]{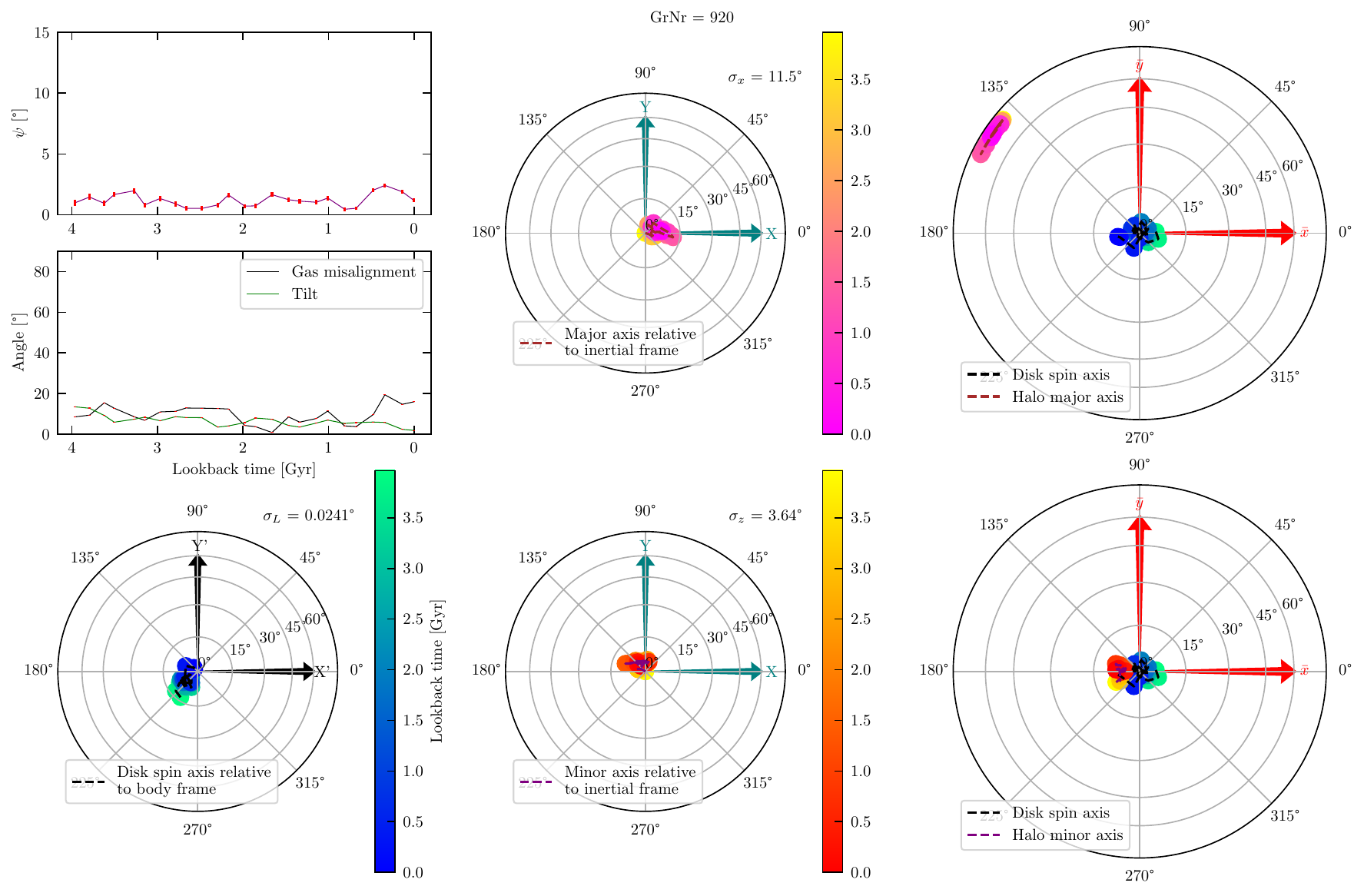}
\caption{Disk-halo evolution for Halo GrNr=920. Top left: Evolution as a function of time of warp angle $\psi$, halo tilt and gas misalignment\mdash Halo 920 shows minimal warping at all times and little variation in both halo tilt and gas misalignment. Bottom left: Orientation of disk spin axis relative to halo body frame as a function of look back time (blue-green color bar) shows little to no variation. Middle column: halo major axis (top) and minor axis (bottom) show minimal evolution in the inertial frame. Right column: disk spin axis (blue-green dots) show little change in orientation but halo major axis shows figure rotation through $\sim 25^\circ$ (top); halo minor axis shows minimal evolution (bottom). Uncertainties on the orientation of each axis ($\sigma$) are given in the top right corner of their respective polar plots. Overall the minimal warping and overall stability of the disk can be attributed to the system's lack of figure rotation and halo tilt.}
\label{fig:920}
\end{figure*}

\begin{figure*}[t]
\centering
\includegraphics[width=0.95\textwidth]{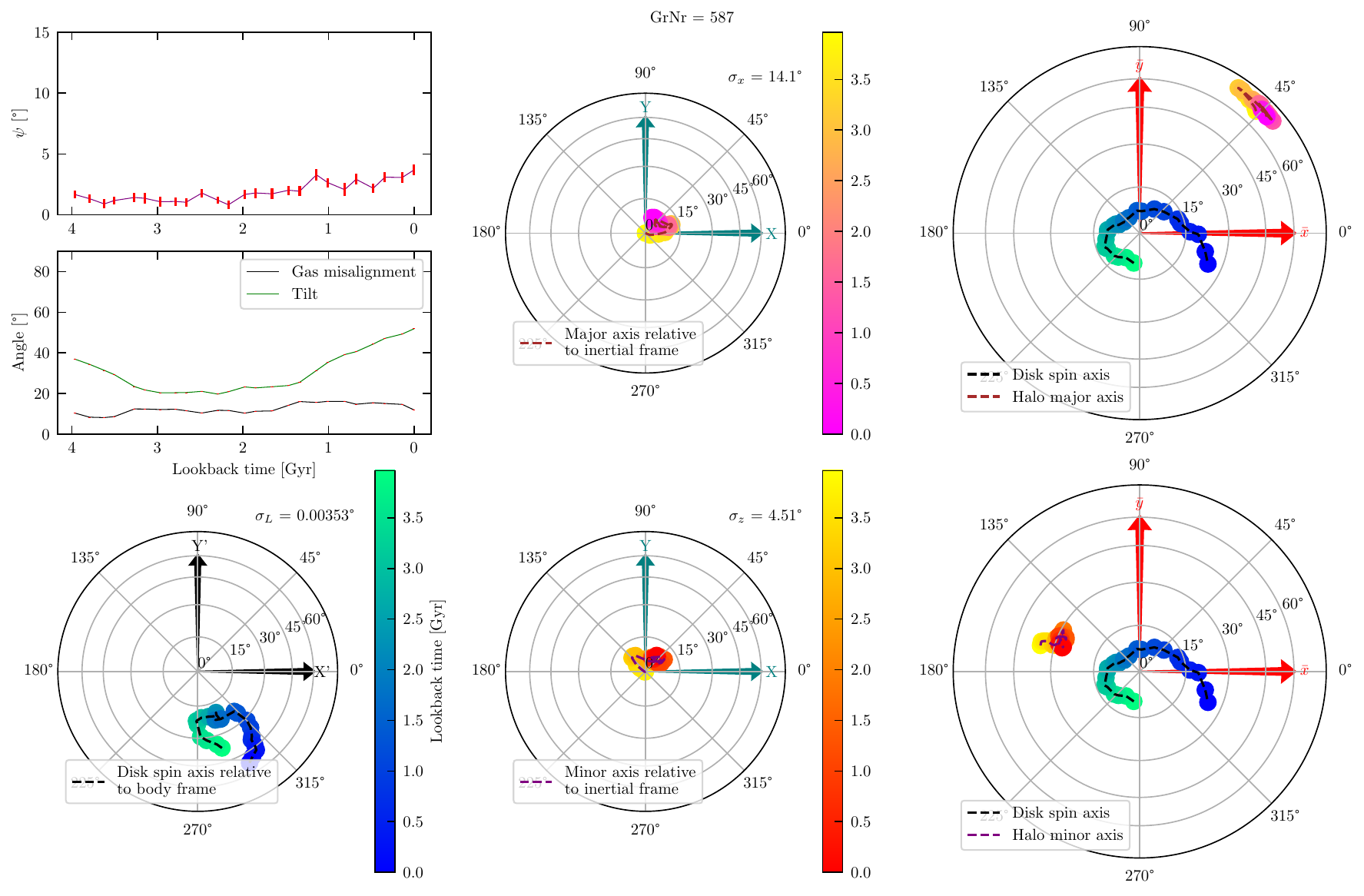}
\caption{The galaxy in Halo 587 exhibits minimal figure rotation (center and right) but has substantial tilt (top right, bottom subplot and bottom right). Since Figure \ref{fig:MCMC} implies correlation of the warp with gas misalignment and halo tilt, we suggest that its steadily growing warp (top right, top subplot) and the disk's slight precession (bottom left, right) may be tied to these factors.}
\label{fig:587}
\end{figure*}

\begin{figure*}
\centering
\includegraphics[width=0.95\textwidth]{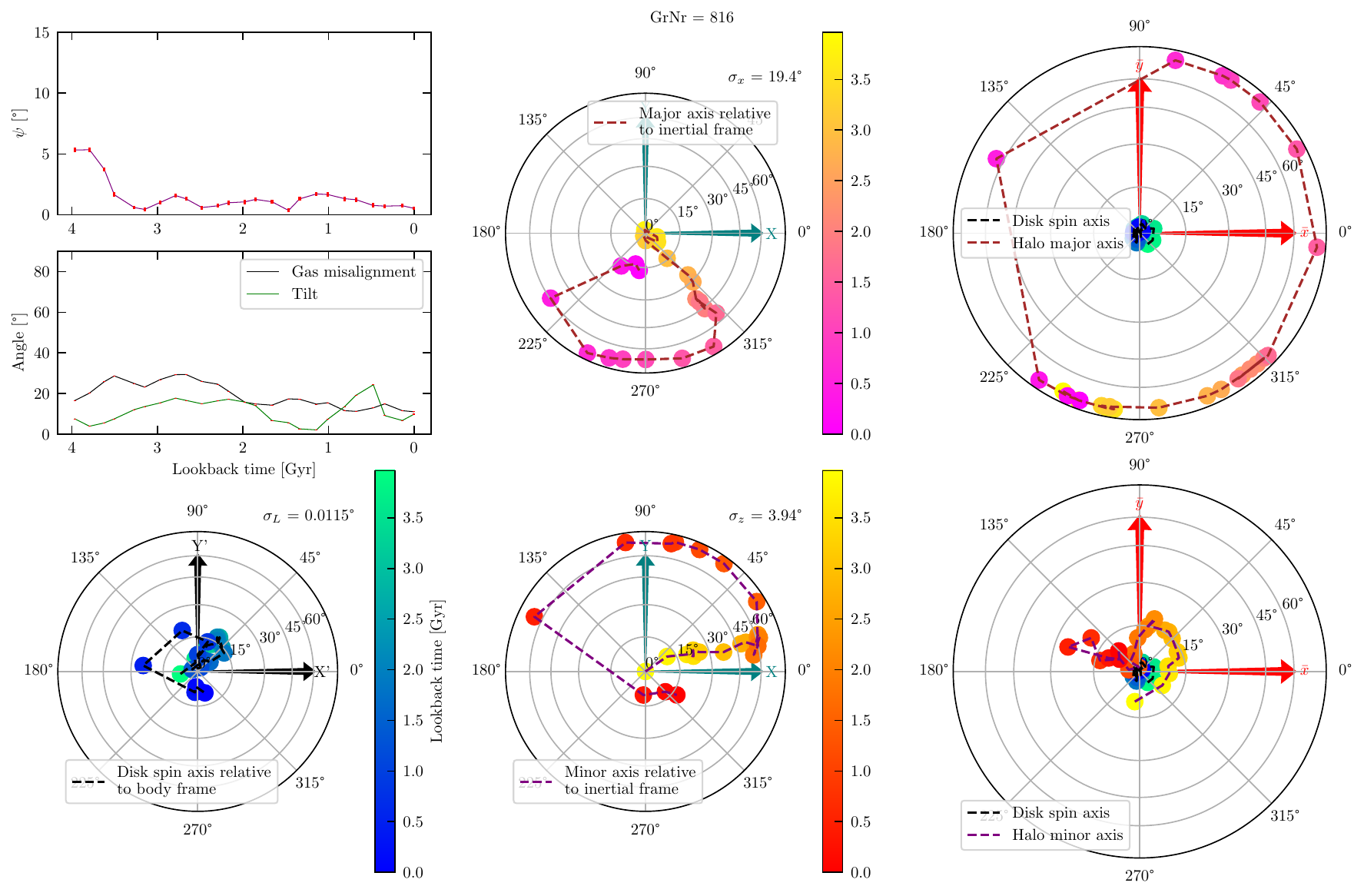}
\caption{The galaxy in Halo 816 exhibits visible figure rotation (center and right), with a similar degree of gas misalignment (top right, bottom subplot) and tilt (top right, bottom subplot and bottom right) compared to Halo 587 in Figure \ref{fig:587}. Figure \ref{fig:MCMC} shows a correlation between both gas misalignment and pattern speed and the warp in this system. It surpasses the population median warp (top left, top subplot) with a halo minor axis that precesses about a mostly stationary disk axis (bottom right), creating the effect of a precessing disk in the non-inertial halo body frame (bottom left). The halo undergoes rather significant figure rotation, as seen by the rapid swinging of the halo major axis with respect to the disk axis.}
\label{fig:816}
\end{figure*}

\begin{figure*}
\centering
\includegraphics[width=1.95\columnwidth]{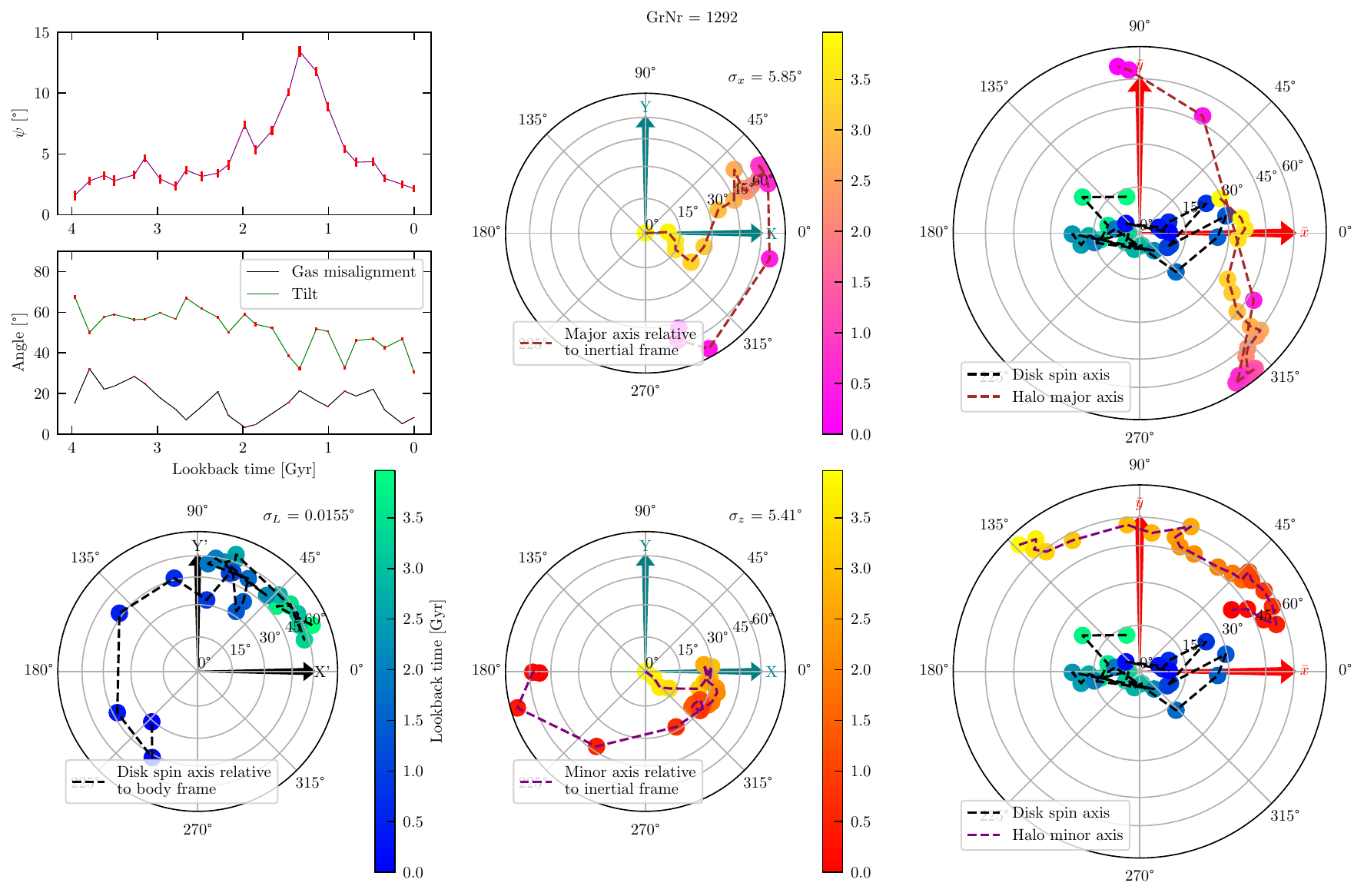}
\caption{The most warped disk in our catalog is Halo 1292, which shows a maximum warp angle approaching $15^\circ$ (top left, top subplot) in the presence of a significantly tilted halo (top left, bottom subplot and right) as well as rapid figure rotation (center and right columns) and disk precession (bottom left and right). The gas is also somewhat misaligned with the disk (bottom subplot in the top left), but not significantly more than in Figure \ref{fig:816}, nor to the optimal warp-causing angle of $45^{\circ}$ found by \cite{Semczuk_2020}. The pattern speed of figure rotation and gas misalignment are shown in Figure \ref{fig:MCMC} to be significant factors in causing the warp.}
\label{fig:1292}
\end{figure*}

To visually track the evolution of the halo figure rotation axis and disk spin axis over time, we employ a variation on Briggs figures \citep{1990ApJ...352...15B}\mdash 2D polar representations of various vector orientations (disk spin axis, halo principal axes). The evolution of each of these four galaxies is illustrated in several ways, as shown in the six panels of Figures~\ref{fig:920}-\ref{fig:1292}. For each of these galaxies, the top left panel shows the evolution of the warp angle $\psi$, gas misalignment angle, and halo tilt angle over time. The 5 polar plots show either the disk or the halo spin axis in the three different frames: the simulation (or ``inertial'') frame (black, $X',Y',Z'$), the ``halo body'' frame (teal $X,Y, Z$ ), and the frame in which the time averaged orientation of the disk angular momentum is the  $z$-axis (red $\bar{x},~ \bar{y}$)\footnote{The $\bar{x},~ \bar{y}$ axes are arbitrarily defined and orthogonal to the averaged orientation of the disk spin axis. We use this frame to more conveniently demonstrate the movement of the disk axis, and stress that this reference frame does not evolve in time.} The three frames are shown in Figure~\ref{fig:frames} with the same three colors used in the polar plots in Figures~\ref{fig:920}-\ref{fig:1292}. In each polar plot, the concentric circles labeled at 15$^\circ$ intervals mark the polar angle relative to the $z$-axis of the respective reference frame (with the corresponding $x, y$ axes indicated by the solid colored arrows). The azimuthal angle in the corresponding equatorial plane is labeled at 45$^\circ$ intervals (anticlockwise from the $x$-axis in each plot). 

The bottom left polar plot shows the orientation of the disk spin axis as a function of lookback time (blue-green color bar) relative to the body frame of the DM halo as shown in \ref{fig:frames}, (black arrows). The middle column shows the orientation over time (colored circles) of the halo major axis (top) and minor axis (bottom) relative to the inertial frame (teal arrows). The halo body frame is rotated such that the axis at the first snapshot (75) the ${z}$-axis lies in the center of the polar plot. In the right hand panels, we center the polar plots on the {\em mean} orientation of the disk spin ($z$) axis over all 25 snapshots (calculated with the covariance matrix of the angular momentum time series) and subsequently plot all axes as a function of lookback time, allowing us to observe the halo and disk spin axis evolutions in their shared context. We note here that it is necessary to use the mean spin axis of the disk since most of the disks wobble or precess over the $\sim 4$~Gyr span that we study them, and because it is nontrivial to define a consistent $x$- and $y$-axis for a tumbling, axisymmetric disk. Uncertainties on the disk spin (angular momentum direction), major, and minor axes orientations ($\sigma_L$, $\sigma_x$, and $\sigma_z$, respectively) are averaged over all 25 snapshots and are annotated in the top right corners of the left and center polar plots.

\paragraph{Halo 920} 
The disk in Halo GrNr 920 is not significantly warped or tilted, and there is minimal misalignment of the gas angular momentum axis with disk spin axis at any time in the last $\sim4$ Gyr (see Figure \ref{fig:920} (top)). In the the bottom left and two right polar plots in Figure \ref{fig:920}, we see that the disk spin axis (blue-green dots) remains at the center of the polar plot, showing that it experiences little to no precession. In the top and center right panels, the halo major ($x$-) axis (yellow-pink dots) can be seen to move a little: between azimuthal angle $\sim 135^\circ-160^\circ$. This is about twice the accuracy with which the position of the major axis can be determined $\sigma_x = 11.5^\circ$, so this is likely a real motion of the major axis. In the bottom right panel, the blue-green points (showing the disk spin axis) and the red-yellow points (showing the orientation of the halo minor axis) mostly overlap the $z$-axis of the coordinate system, showing that the disk spin axis is aligned with the halo minor axis over the entire duration examined. The overall inference from Figure \ref{fig:920} is that the galaxy in Halo 920 has a stable disk, showing little evidence of warping and tilting. Its halo minor axis is always aligned with the disk spin axis and shows minimal, steady figure rotation. We include this galaxy as an example of a system that is relatively unperturbed with comparably little dynamical evolution, and refer to it as our ``control'' halo.

\paragraph{Halo 587}
The disk in Halo GrNr 587 (Figure \ref{fig:587} (top left, top subplot)) is more warped than the disk in GrNr 920. In the two right and the bottom left polar plots of Figure \ref{fig:587}, we see that the disk spin axis (blue-green dots) precesses along an arc $\sim30^\circ$ in diameter and sweeps out $\sim310^\circ$ in azimuth, nearly completing a full period in the last 4 Gyr. We can see in the bottom right panel and the top left panel's bottom subplot that the tilt between the disk spin axis and halo minor axis is significantly larger than for GrNr 920; the blue-green points (showing the disk spin axis) and the red-yellow points (showing the orientation of the halo minor axis) are between $\sim 15^\circ-60^\circ$ apart. The rather consistent offset between the halo minor axis and the center of the arc swept out by the disk axis suggests that GrNr 587 may have a twisted DM halo, whose minor axis in the inner regions is misaligned with the minor axis of the outer halo. The top left panel's bottom subplot also shows slight gas misalignment. GrNr 587 is therefore another halo with minimal figure rotation, although it is more warped than our control case. Figure \ref{fig:MCMC} shows that gas misalignment and tilt are both correlated significantly with disk warping in this system. It is likely that both the gas misalignment and halo tilt in this system contribute to the observed disk precession and warping. Among our four case studies, this system is the most similar to that studied by \cite{han2023tilteddarkhaloscommon}.

\paragraph{Halo 816}
In the two rightmost panels of Figure \ref{fig:816}, we see that the disk spin axis (blue-green dots) remains rather stationary in the inertial frame while the halo minor ($z$-) axis (red-yellow dots) precesses about it, creating the effect of disk precession in the halo body frame (bottom left). The halo minor axis precesses about an arc $\sim15^\circ$ in diameter until $\sim1.5$ Gyr ago, while the halo major axis demonstrates clear figure rotation, sweeping through $\gtrsim360^\circ$ over the time series. The halo is also clearly tilted, as seen in the lower of the two subplots in the top left and the separation between the disk spin axis (blue-green dots) and halo minor axis (red-yellow dots) in the bottom right panel. It also has a similar degree of gas misalignment as Halo 587 (top left, bottom subplot). Figure \ref{fig:MCMC} shows that gas misalignment and figure rotation pattern speed are both significantly correlated with disk warping in this galaxy. It is likely that misaligned gas accretion and halo figure rotation are the most important contributors to the warp in this system. We also note that since the highest warp we observe in this system occurs at the beginning of the time series, given the lags our analysis is sensitive to, we cannot suggest a likely cause of this warp. Rather, what we measure are correlations with the relatively smaller warps occurring later in the time series.

\paragraph{Halo 1292}
Our final case study is GrNr 1292. Figure \ref{fig:1292} shows the most warped disk (top left, top subplot) in our catalog (the galaxy shown in Figure \ref{fig:warp fitting}). In the center and right panels, the halo major ($x$-) and minor ($z$-) axes (pink-yellow dots and red-yellow dots) display a significant motion over the sampled time course. In the time-averaged disk frame (right column), both the halo minor and major axis are seen to swing about wide angles in azimuth, subtending a range of $\sim150^\circ$ for the major axis and $\sim120^\circ$ for the minor axis. The apparent motion of both axes suggests that this halo is undergoing figure rotation about an axis misaligned with any principal axis but nearly aligned with the mean disk axis. The disk spin axis (blue-green) in the halo body frame (bottom left) shows a similarly large degree of precession. As greater precession in the presence of a more tilted halo (top left, bottom subplot and right) occurs, this disk achieves the highest warp in our catalog. While there is a greater misalignment with the gas than in the first two case studies seen in Figures \ref{fig:920} and \ref{fig:587} (top left panel, bottom left subplots), our MC analysis in Figure \ref{fig:MCMC} suggests that both gas misalignment and the figure rotation observed in Figure \ref{fig:1292} are significantly correlated with disk warping in this system, and therefore each may be important contributors to forming the observed warp at $t_{lb}\sim1.2$ Gyr.

By quantifying the range of warps achieved by the stellar disks in these four case studies, we can dissect the extent to which increased warping coincides with the presence of figure rotation, halo tilt and gas misalignment. Figure \ref{fig:587} shows that even in isolated galaxies, moderate warping can occur in the presence of halo tilt. Figures \ref{fig:816} and \ref{fig:1292} show that higher warping can arise in the presence of higher degrees of gas misalignment and figure rotation. Thus, we see that disk warps coincide with a variety of dynamical considerations: a tilted halo's torque on the disk, a tumbling halo exerting a torque or Coriolis force on the disk, a halo and disk moving in paired precessing motions, then finally a gas angular momentum misaligned with the disk's spin axis are all able to contribute to the warps we observe in our galaxies. 

\section{Discussion and Conclusion} \label{sec: summary}


Recent studies of the Milky Way and TNG50 disk galaxies have shown that stellar and DM halos may commonly be tilted relative to galactic disks \citep{Han2022tilted_stellar}, and such tilted halos can cause warps \citep{Han2023NatAst_warp, han2023tilteddarkhaloscommon}. Disk warping can develop within 1 Gyr following the formation of a halo tilt \citep{han2023tilteddarkhaloscommon}, and may begin to develop in as little as one circular period ($\sim400$ Myr) \citep{Han2023NatAst_warp}. Both halo tilt (misalignment) and halo figure rotation are expected to arise from past major mergers \citep[][Ash et al. 2026 in prep.]{arora_shaping_25} and are closely related. In particular, if a halo's principal axis is misaligned with the disk spin axis, previous works have shown that the disk will realign itself to be close to a principal axis of the halo on a fairly short timescale without significantly changing shape \citep{Debattista2013_whatsup}\mdash implying that it will undergo disk tilting in the process. AV23 also showed that figure rotation is ubiquitous, stable and long-lived in isolated halos and Coriolis forces from a tumbling halo may also be capable of driving warping in isolated disks \citep{Dubinski_2009}. 

In this work, we have investigated the generation of warps in a sample of TNG50 disk galaxies to assess whether halo tilting, halo figure rotation, or gas misalignment can be considered the primary driver of warps in isolated disks. A primary goal for this work was to determine whether warps in isolated disk galaxies can be used as a probe of halo figure rotation. To this end, we analyzed warps in 40 disk galaxies from the TNG50 suite that were relatively isolated over the last $\sim$ 4 Gyr and belong to the halo catalog studied by AV23.  We  calculated the warp angles $\psi$ using an azimuthal harmonic decomposition method and linear fitting in our disks over the last $\sim4$ Gyr (25 snapshots), finding a median $\psi_{\max}$ over all galaxies of $5.4^\circ$ and 75$^{\text{th}}$ percentile of $6.7^\circ$ (see Figure \ref{fig:kde}).

To assess whether they may be significant drivers of later warping, we performed a modified cross-correlation analysis between 4 potential warp drivers (halo tilt, gas angular momentum misalignment, figure rotation pattern speed, and figure rotation axis misalignment) and disk warping after some time lag between $0-2.3$ Gyr. We did not find that any single lag, nor any narrow distribution of lags, yielded a significant correlation between any of our warp drivers and warp angle at the population level, even when we control for the varying dynamical times of the disks in our sample. This lack of a clearly favored lag or significant averaged correlation is an indication that no individual warp driver is dominant in our sample for all, or even most, galaxies. Our attempts to find a common timescale on which a given driver can excite disk warping were likely unsuccessful because the warps in our sample are driven by the combination of many simultaneous torques, making it difficult to identify a clear signal. Future attempts to measure these timescales could possibly be improved with either higher time resolution outputs, allowing us to probe a finer grid of time lags, or with a larger population of cosmological disks.

While our population-averaged correlation coefficients were low, individual galaxies in our sample did achieve significant correlations. To better determine the significance of our cross-correlation results, we focus on the maximum Pearson correlation coefficient ($r_{\mathrm{max}}$) achieved at any lag, for each galaxy. By performing a Monte Carlo experiment, we found that the maximum correlation coefficients achieved by our cross-correlation analysis were significantly offset from those expected under random chance for all four of our considered warp drivers, suggesting that each may be somewhat related to later warping in the average galaxy within our sample. However, no individual driver was significant in all systems, or was uniformly more important than the other drivers. Figure rotation pattern speed and gas misalignment showed the greatest offsets, suggesting that these typically show the strongest correlation with disk warping at some later time. In contrast, the $r_{\mathrm{max}}$ achieved by the correlations between misalignment of the figure rotation axis and subsequent warping was nearly consistent with our Monte Carlo distributions representing random chance, suggesting that figure rotation misalignment was not a particularly important driver of disk warping in our sample.

In a detailed examination of individual systems, we have shown that both halo tilt and rapid figure rotation can coincide with the presence of warps. In particular, we found that warped disks can arise in galaxies with tilted halos (see Figure \ref{fig:587}); disks can also exhibit precession paired with the movement of halo principal axes, which together may contribute to warping (see Figure \ref{fig:816}); and finally, the most warped disks we observe occur in halos with significant figure rotation (see Figure \ref{fig:1292}). We stress again that for every warp greater than 7$^\circ$ that occurred in our sample, the halo had a significant degree of figure rotation but not necessarily gas misalignment. 

We emphasize that within this study we only measure correlations between known warp drivers and subsequent warping, and we do not attempt to determine whether the warp is causally related with a given driver. We do this for two reasons: 
\begin{enumerate}
    \item Each of the warp drivers we consider (with the exception of figure rotation axis misalignment) have previously been shown to be capable of generating warps in isolation, and 
    \item In a cosmological environment, many time-dependent confounding factors\mdash such as orbiting substructures and torques from large-scale structure\mdash are present, making a demonstration of causality restrictively burdensome.
\end{enumerate}

Instead of demonstrating causality, we have considered the strengths of correlations as evidence (or lack thereof) for the relative importance of various warp drivers in a cosmological setting. A potential weakness of this approach arises if the warp drivers we consider are strongly correlated with one another, or if warping and any of our drivers are simultaneously produced by a perturbation we have not considered. We mitigate the risks of this kind of spurious correlation by considering primarily population statistics (either $\bar{r}$ or $r_{\mathrm{max}}$ distributions) which are less likely to be impacted.

While we do not consider it as a warp driver in this work, the torques from a tilting or precessing inner disk can also be an important driver of warping in the outer disk. Indeed we do observe precessing disks within our sample (see e.g. Figure \ref{fig:587}). Disk precession of this nature is typically driven by external torques exerted on the inner disk, such as may be caused by a tilted halo. A tilted halo is indeed present for Halo 587 shown in Figure \ref{fig:587}, as the disk is consistently misaligned with the minor axis of the halo and does not realign with this minor axis but instead precesses about another point. Since our measurement of the minor axis is made in the outer halo, this scenario is likely explained either by a twisted halo (i.e., the minor axis in the outer halo is not aligned to the minor axis of the inner halo) or by a spherical inner halo with a tilted external halo, as was studied by \cite{Han2023NatAst_warp}. Although in these cases the warp will be driven by the vector summation of the torques on the outer disk by the tilting inner disk and by the tilted DM halo (in the absence of any other torques), we investigate only halo tilt as a driver of warping because the tilting of the inner disk is also caused by the halo tilt.

\citet{Earp_2019_disktilt} found from an examination of tilting disks in an isolated sample of galaxies from the NIHAO simulations \citep{2015MNRAS.454...83W} that there is a strong correlation between the tilting rate of the stellar disk and the misalignment of the warp in the cold gas component. They also find that none of the DM halo's properties that they examined appear to be correlated with the tilting, and conclude that gas cooling on to the disk is the principal driver of disk tilting. In this work we have not investigated the effects of cold gas accretion on the warping of disks since this is an issue that has been studied extensively in the past \citep{1989MNRAS.237..785O,2008A&ARv..15..189S,2010MNRAS.408..783R,2017MNRAS.465.3446G,2018MNRAS.474..254R}. Instead, our focus has been restricted to examining the effects of halo-disk misalignment and halo figure rotation, and we attempt to control for the accretion of cold gas by considering the misalignment between the disk angular momentum and the gas angular momentum outside and around the disk.

Our results show that warps are often formed in the presence of halo figure rotation. Besides figure rotation, we find several other factors that can cause warps, either simultaneously in the same galaxy or independently in a galaxy without figure rotation. In agreement with previous work, we conclude that warps can be excited by tidal interactions, cold gas accretions, and halo-disk tilting. Our study suggests that in order for warps in external galaxies to serve as tracers of figure rotation in their dark halos, very careful analysis and possibly additional dynamical tracers (such as streams in their stellar halos) would be required to rule out alternative causes. In particular, future work is required to determine what tracers may be required to distinguish warps caused by tilted DM halos and warps caused by halo figure rotation. Nevertheless, our study shows that despite the fact that the pattern speeds of halo figure rotation ($\sim 10^{\circ}$/Gyr) are small, they play a far from insignificant role on the structure and dynamical evolution of the disk galaxies that reside in them.

\section{Acknowledgments} \label{sec: thanks} SJ thanks Drew Lapeer for a friendship and support that led to the pursuit of galactic dynamics and Michelle Jecmen for her contribution to the early stages of this project. We also thank the authors of \cite{refId0}, particularly Ilia V. Chugunov, for access to their warp data (shown in Figure~\ref{fig:kde}). We gratefully acknowledge funding from NASA-ATP awards 80NSSC20K0509 and 80NSSC24K0938 to MV.

\section{Author contributions} S.~Johri carried out the majority of the analysis of the disk galaxies, measured the properties of warped galaxies, and wrote the paper. N.F.~Ash carried out the analysis of DM halo shapes, figure rotation, formulated the algorithm for modeling the warp, and assisted with analysis and writing, including the construction and testing of Monte Carlo distributions. M.~Valluri was responsible for overall intellectual guidance and edited the paper. All authors contributed to the final manuscript and approved it for submission.


\software{Agama \citep{2019MNRAS.482.1525V}, NumPy \citep{harris2020array}, SciPy\_python \citep{Oliphant2007}.}

\bibliographystyle{aasjournal}
\bibliography{references}

\end{document}